\documentstyle[amssymb,11pt]{article}

\newlength{\minitwocolumn}
\font\teneufm=eufm10
\font\seveneufm=eufm7
\font\fiveeufm=eufm5
\newfam\eufmfam
\textfont\eufmfam=\teneufm
\scriptfont\eufmfam=\seveneufm
\scriptscriptfont\eufmfam=\fiveeufm

\makeatletter
\@addtoreset{equation}{section}
\makeatother

\title{\bf
\huge{\bf
Infinitely many commuting operators 
for
the elliptic quantum group $U_{q,p}(\widehat{sl_N})$}
}
\begin{document}
\maketitle
\begin{center}
{Takeo KOJIMA}
\\~\\
{\it
Department of Mathematics and Physics,
Graduate School of Science and Engineering,\\
Yamagata University, Jonan 4-3-16, Yonezawa 992-8510,
Japan}
\end{center}
~\\
\begin{abstract}
We construct
two classes of infinitely many commuting operators
associated with the elliptic quantum group 
$U_{q,p}(\widehat{sl_N})$.
We call one of them 
the integral of motion ${\cal G}_m$,
$(m \in {\mathbb N})$ and the other
the boundary transfer matrix 
$T_B(z)$, $(z \in {\mathbb C})$.
The integral of motion ${\cal G}_m$ is related
to elliptic deformation of the $N$-th KdV theory.
The boundary transfer matrix $T_B(z)$ is related to
the boundary $U_{q,p}(\widehat{sl_N})$ face model.
We diagonalize the boundary transfer matrix
$T_B(z)$ by 
using the free field realization of 
the elliptic quantum group,
however diagonalization of the integral of motion ${\cal G}_m$
is open problem even for the simplest case $U_{q,p}(\widehat{sl_2})$.
\end{abstract}

~\\
\section{Introduction}

The free field approach provides 
a powerful method to study exactly solvable model
\cite{JM}.
The basic idea in this approach
is to realize the commutation relations for the symmetry
algebra and the vertex operators in terms of
free fields acting on the Fock space.
%In this paper
%we are interested in
%the elliptic quantum group $U_{q,p}(\widehat{sl_N})$ that
%is elliptic generalization of the quantum group 
%$U_q(\widehat{sl_N})$.
We introduce the elliptic quantum group
$U_{q,p}(\widehat{sl_N})$ \cite{JKOS, KojimaKonno},
and give its free field realization.
Using the free field realizations,
we introduce two extended currents
$F_N(z)$ \cite{KojimaShiraishi1} and 
$U(z)$ \cite{AJMP} associated with 
the elliptic quantum group $U_{q,p}(\widehat{sl_N})$.
We construct
two classes of infinitely many commuting operators
for the elliptic quantum group
$U_{q,p}(\widehat{sl_N})$.
We call one of them the integral of motion ${\cal G}_m$,
$(m \in {\mathbb N})$ \cite{KojimaShiraishi1}
and the other
the boundary transfer matrix 
$T_B(z)$, $(z \in {\mathbb C})$ \cite{Kojima3}.
Our constructions are based on
the free field realizations of
the elliptic quantum group 
$U_{q,p}(\widehat{sl_N})$,
the extended currents and the vertex operator
$\Phi^{(a,b)}(z)$.
Commutativity of 
the integral of motion is ensured
by Feigin-Odesskii algebra \cite{FO},
and those of
the boundary transfer matrix
is ensured by Yang-Baxter equation and 
boundary Yang-Baxter equation \cite{S}.
Two classes of infinitely many commuting operators
have physical meanings.
The integral of motion ${\cal G}_m$ is 
two parameter deformation of
the monodromy of the $N$-th KdV theory
\cite{BLZ, Kojima2}.
The boundary transfer matrix $T_B(z)$ is related to
the boundary $U_{q,p}(\widehat{sl_N})$ face model
that is lattice deformation of 
the conformal field theory.
We diagonalize the boundary transfer matrix
$T_B(z)$ by using the free field realization of 
the elliptic quantum group and the vertex operators.
Diagonalization of the boundary transfer matrix 
allows us calculate correlation functions
of the boundary $U_{q,p}(\widehat{sl_N})$ face model
\cite{LP, MW, Kojima3}.

The organization of this paper is as follows.
In section 2 we introduce
the elliptic quantum group
$U_{q,p}(\widehat{sl_N})$ \cite{JKOS, KojimaKonno},
and give its free field realization.
In section 3
we introduce two extended currents
$F_N(z),E_N(z)$ \cite{KojimaShiraishi1}
and $U(z),V(z)$ \cite{AJMP, FKQ}
associated with 
the elliptic quantum group $U_{q,p}(\widehat{sl_N})$.
We give the free field realization of the vertex operators
$\Phi^{(a,b)}(z)$, using the extended current $U(z)$.
We construct
two classes of infinitely many commuting operators
associated with 
the elliptic quantum group
$U_{q,p}(\widehat{sl_N})$.
The one is the integral of motion ${\cal G}_m$
\cite{KojimaShiraishi1}
and the other is 
the boundary transfer matrix $T_B(z)$ \cite{Kojima3}.
In section 4 we diagonalize the boundary transfer matrix
$T_B(z)$ by 
using the free field realization of 
the vertex operators
\cite{AJMP, FKQ, Kojima3}.

\section{Elliptic quantum group $U_{q,p}(\widehat{sl_N})$}
 
In this section we introduce the elliptic quantum group 
$U_{q,p}(\widehat{sl_N})$ and its free field realization.

\subsection{Quantum group}

In this section
we recall Drinfeld realization of the quantum group
\cite{Drinfeld}.
We fix a complex number $q$ such that $0<|q|<1$.
Let us fix the integer $N=3,4,5,\cdots$.
We use q-integer $[n]_q=\frac{q^a-q^{-a}}{q-q^{-1}}$.
We use the abbreviation,
\begin{eqnarray}
(z;p_1,p_2,\cdots,p_M)_\infty
=\prod_{k_1,k_2,\cdots,k_M=0}^\infty
(1-p_1^{k_1}p_2^{k_2}\cdots p_M^{k_M}z).\nonumber
\end{eqnarray}
The quantum group $U_q(\widehat{sl_N})$
is generated by $h_j, a_{j,m}, x_{j,n}$,
$(1\leqq j \leqq N-1:m\in {\mathbb Z}_{\neq 0},n \in
{\mathbb Z})$, $c,d$.
Let us set the generating functions 
$x_j^\pm(z), \psi_j(z), \varphi_j(z)$, 
$(1\leqq j \leqq N-1)$ by
\begin{eqnarray}
x_j^\pm(z)&=&
\sum_{n \in {\mathbb Z}}x_{j,n}^\pm z^{-n},\nonumber
\\
\psi_j(q^{\frac{c}{2}}z)&=&q^{h_j}
\exp\left((q-q^{-1})\sum_{m>0}a_{j,m}z^{-m}\right),
\nonumber
\\
\varphi_j(q^{-\frac{c}{2}}z)&=&
q^{-h_j}
\exp\left(-(q-q^{-1})\sum_{m>0}a_{j,-m}z^{m}\right).
\nonumber
\end{eqnarray}
The defining relations are given by
\begin{eqnarray}
&&~[d,x_{j,n}^\pm]=n x_{j,n}^\pm,
~[h_j,d]=[h_j,a_{k,m}]=[d,a_{k,m}]=0,
c : {\rm central},
\nonumber\\
&&
~[a_{j,m},a_{k,n}]=\frac{[A_{j,k}m]_q [cm]_q}{m}
q^{-c|m|}\delta_{m+n,0},
~[h_j,x_k^\pm(z)]=\pm A_{j,k} x_k^\pm(z),\nonumber\\
&&
~[a_{j,m},x_k^+(z)]=\frac{[A_{j,k}m]_q}{m}q^{-c|m|}z^m 
x_k^+(z),
~[a_{j,m},x_k^-(z)]=-\frac{[A_{j,k}m]_q}{m}z^m x_k^-(z),
\nonumber\\
&&
(z_1-q^{\pm A_{j,k}}z_2)x_j^\pm(z_1)x_k^\pm(z_2)=
(q^{\pm A_{j,k}}z_1-z_2)x_k^\pm(z_2)x_j^\pm(z_1),\nonumber\\
&&~[x_j^+(z_1),x_k^-(z_2)]=\frac{\delta_{j,k}}{q-q^{-1}}
(\delta(q^{-c}z_1/z_2)\psi_j(q^{\frac{c}{2}}z_2)-
\delta(q^cz_1/z_2)\varphi_j(q^{-\frac{c}{2}}z_2)),\nonumber
\end{eqnarray}
and Serre relation for $|j-k|=1$,
\begin{eqnarray}
&&(x_j^\pm(z_1)x_j^\pm(z_2)x_k^\pm(z)-
[2]_q
x_j^\pm(z_1)x_k^\pm(z)x_j^\pm(z_2)
+x_k^\pm(z)x_j^\pm(z_1)x_j^\pm(z_2))\nonumber\\
&&+
(x_j^\pm(z_2)x_j^\pm(z_1)x_k^\pm(z)-
[2]_q
x_j^\pm(z_2)x_k^\pm(z)x_j^\pm(z_1)
+x_k^\pm(z)x_j^\pm(z_2)x_j^\pm(z_1))=0.\nonumber
\end{eqnarray}
Here $(A_{j,k})_{1\leqq j,k \leqq N-1}$
is Cartan matrix of $sl_N$ type.
Here we used the delta function 
$\delta(z)=\sum_{m \in {\mathbb Z}}z^m$.

\subsection{Elliptic quantum group}

In this section
we introduce the elliptic quantum group
$U_{q,p}(\widehat{sl_N})$
\cite{JKOS, KojimaKonno}, which is
elliptic deformation of 
the quantum group $U_q(\widehat{sl_N})$.
We fix complex numbers $r, s$ such that
${\rm Re}(r)>1$ and ${\rm Re}(s)>0$.
When we change the polynomial
$(z_1-q^{-2}z_2)$ in the defining relation
of the quantum group $U_q(\widehat{sl_N})$,
\begin{eqnarray}
(z_1-q^{-2}z_2)x_j^-(z_1)x_j^-(z_2)=
(q^{-2}z_1-z_2)x_j^-(z_2)x_j^-(z_1),\nonumber
\end{eqnarray}
to the elliptic theta function 
$[u]$, we have
\begin{eqnarray}
~\left[u_1-u_2+1\right] F_j(z_1)F_j(z_2)
=\left[u_1-u_2-1\right] F_j(z_2)F_j(z_1).
\nonumber
\end{eqnarray}
This is one of the defining relations of 
the elliptic quantum
group $U_{q.p}(\widehat{sl_N})$.
We set the elliptic theta function $[u], [u]^*$ by
\begin{eqnarray}
&&[u]=q^{\frac{u^2}{r}-u}\Theta_{q^{2r}}(q^{2u}),
~~[u]^*=q^{\frac{u^2}{r^*}-u}\Theta_{q^{2r^*}}(q^{2u}),
\nonumber
\\
&&\Theta_p(z)=(p;p)_\infty (z;p)_\infty (pz^{-1};p)_\infty,
\nonumber
\end{eqnarray}
where we set $z=x^{2u}$ and $r^*=r-c$.
The elliptic quantum group $U_{q,p}(\widehat{sl_N})$
is generated by the currents
$E_j(z), F_j(z)$, 
$H_j^+(q^{\frac{c}{2}-r}z)=H_j^-(q^{-\frac{c}{2}+r}z)$, 
$(1\leqq j \leqq N-1)$.
The defining relations are given by
\begin{eqnarray}
E_j(z_1){E}_{j+1}(z_2)
&=&
\frac{
\left[u_2-u_1+\frac{s}{N}\right]^*}
{\left[u_1-u_2+1-\frac{s}{N}\right]^*}
{E}_{j+1}(z_2){E}_j(z_1),\\
{E}_j(z_1)
{E}_j(z_2)&=&
\frac{[u_1-u_2+1]^*}
{[u_1-u_2-1]^*}
{E}_j(z_2){E}_j(z_1),\\
{E}_j(z_1){E}_k(z_2)&=&{E}_k(z_2){E}_j(z_1),~~
otherwise ,
\end{eqnarray}
\begin{eqnarray}
{F}_j(z_1){F}_{j+1}(z_2)
&=&\frac{\left[u_2-u_1+\frac{s}{N}-1\right]}{
\left[u_1-u_2-\frac{s}{N}\right]
}{F}_{j+1}(z_2){F}_j(z_1),\\
{F}_j(z_1){F}_j(z_2)&=&
\frac{[u_1-u_2-1]}{[u_1-u_2+1]
}{F}_j(z_2){F}_j(z_1),\\
{F}_j(z_1){F}_k(z_2)&=&
{F}_k(z_2){F}_j(z_1),~~
otherwise ,
\end{eqnarray}
\begin{eqnarray}
H_j^+(z_1)H_j^+(z_2)
&=&
\frac{[u_1-u_2-1]
[u_1-u_2+1]^*}{
[u_1-u_2+1]
[u_1-u_2-1]^*}
H_j^+(z_2)H_j^+(z_1),\\
H_j^+(z_1)H_{j+1}^+(z_2)
&=&
\frac{[u_1-u_2+1-\frac{s}{N}]
[u_1-u_2-\frac{s}{N}]^*}{
[u_1-u_2-\frac{s}{N}]
[u_1-u_2+1-\frac{s}{N}]^*}
H_{j+1}^+(z_2)H_j^+(z_1),\\
H_j^+(z_1)H_k^+(z_2)
&=&H_k^+(z_2)H_j^+(z_1),~
otherwise,
\end{eqnarray}
\begin{eqnarray}
H_j^+(z_1)E_j(z_2)&=&
\frac{[u_1-u_2+1+\frac{c}{4}]^*}
{[u_1-u_2-1-\frac{c}{4}]^*}
E_j(z_2)H_j^+(z_1),\\
%%%%%%%%%%%%%%%%%%%%%%%%%%%
H_j^+(z_1)E_{j+1}(z_2)&=&
\frac{[u_2-u_1+\frac{s}{N}+\frac{c}{4}]^*}
{[u_1-u_2+1-\frac{s}{N}-\frac{c}{4}]^*}
E_{j+1}(z_2)H_j^+(z_1),\\
%%%%%%%%%%%%%%%%%%%%%%%%%%%
H_{j+1}^+(z_1)E_j(z_2)&=&
\frac{[u_2-u_1+1-\frac{s}{N}+\frac{c}{4}]^*}
{[u_1-u_2-\frac{s}{N}-\frac{c}{4}]^*}
E_j(z_2)H_{j+1}^+(z_1),\\
H_{j}^+(z_1)E_k(z_2)&=&
E_k(z_2)H_{j}^+(z_1),~otherwise,
\end{eqnarray}
%%%%%%%%%%%%%%%%%%%%%%%%%%%
\begin{eqnarray}
H_j^+(z_1)F_j(z_2)&=&
\frac{[u_1-u_2-1-\frac{c}{4}]}
{[u_1-u_2+1+\frac{c}{4}]}
F_j(z_2)H_j^+(z_1),\\
%%%%%%%%%%%%%%%%%%%%%%%%%%%
H_j^+(z_1)F_{j+1}(z_2)&=&
\frac{[u_2-u_1+\frac{s}{N}-1-\frac{c}{4}]}
{[u_1-u_2-\frac{s}{N}+\frac{c}{4}]}
F_{j+1}(z_2)H_j^+(z_1),\\
%%%%%%%%%%%%%%%%%%%%%%%%%%%
H_{j+1}^+(z_1)F_j(z_2)&=&
\frac{[u_2-u_1-\frac{s}{N}-\frac{c}{4}]}
{[u_1-u_2+\frac{s}{N}-1+\frac{c}{4}]}
F_j(z_2)H_{j+1}^+(z_1),\\
H_j^+(z_1)F_k(z_2)&=&F_k(z_2)H_j^+(z_1),~otherwise,
\end{eqnarray}
%%%%%%%%%%%%%%%%%%%%%%%%%%%
\begin{eqnarray}
~[E_i(z_1),F_j(z_2)]&=&\frac{\delta_{i,j}}{q-q^{-1}}
\left(\delta(q^{-c}z_1/z_2)
H_j^+\left(q^{\frac{c}{2}}z_2\right)
-\delta(q^{c}z_1/z_2)
H_j^-\left(q^{-\frac{c}{2}}z_2\right)\right),\nonumber\\
\end{eqnarray}
and the Serre relations for $|j-k|=1$,
\begin{eqnarray}
&&\left\{
(z_2/z)^{\frac{1}{r^*}}
\frac{
(q^{2r^*-1}z/z_1;q^{2r^*})_\infty
(q^{2r^*-1}z/z_2;q^{2r^*})_\infty}{
(q^{2r^*+1}z/z_1;q^{2r^*})_\infty 
(q^{2r^*+1}z/z_2;q^{2r^*})_\infty}
E_j(q^{1-\frac{2s}{N}}z_1)
E_j(q^{1-\frac{2s}{N}}z_2)
E_k(q^{1-\frac{2s}{N}}z)
\right.\nonumber
\\
&&-[2]_q
\frac{(q^{2r^*-1}z/z_1;q^{2r^*})_\infty
(q^{2r^*-1}z_2/z;q^{2r^*})_\infty}{
(q^{2r^*+1}z/z_1;q^{2r^*})_\infty 
(q^{2r^*+1}z_2/z;q^{2r^*})_\infty}
E_j(q^{1-\frac{2s}{N}}z_1)
E_k(q^{1-\frac{2s}{N}}z)
E_j(q^{1-\frac{2s}{N}}z_2)\nonumber
\\
&&\left.
+(z/z_1)^{\frac{1}{r^*}}
\frac{(q^{2r^*-1}z_1/z;q^{2r^*})_\infty
(q^{2r^*-1}z_2/z;q^{2r^*})_\infty}{
(q^{2r^*+1}z_1/z;q^{2r^*})_\infty 
(q^{2r^*+1}z_2/z;q^{2r^*})_\infty}
E_k(q^{1-\frac{2s}{N}}z)
E_j(q^{1-\frac{2s}{N}}z_1)
E_j(q^{1-\frac{2s}{N}}z_2)
\right\}\nonumber\\
&\times&
z_1^{-\frac{1}{r^*}}
\frac{(q^{2r^*+2}z_2/z_1;q^{2r^*})_\infty}
{(q^{2r^*-2}z_2/z_1;q^{2r^*})_\infty}
+(z_1\leftrightarrow z_2)=0,\\
%%%%%%%%%%%%%%%%%%%%%%%%%%%%%%%%
&&\left\{
(z_2/z)^{-\frac{1}{r}}
\frac{
(q^{2r+1}z/z_1;q^{2r})_\infty
(q^{2r+1}z/z_2;q^{2r})_\infty}{
(q^{2r-1}z/z_1;q^{2r})_\infty 
(q^{2r-1}z/z_2;q^{2r})_\infty}
F_j(q^{1-\frac{2s}{N}}z_1)
F_j(q^{1-\frac{2s}{N}}z_2)
F_k(q^{1-\frac{2s}{N}}z)
\right.\nonumber
\\
&&-[2]_q
\frac{(q^{2r+1}z/z_1;q^{2r})_\infty
(q^{2r+1}z_2/z;q^{2r})_\infty}{
(q^{2r-1}z/z_1;q^{2r})_\infty 
(q^{2r-1}z_2/z;q^{2r})_\infty}
F_j(q^{1-\frac{2s}{N}}z_1)
F_k(q^{1-\frac{2s}{N}}z)
F_j(q^{1-\frac{2s}{N}}z_2)\nonumber
\\
&&\left.
+(z/z_1)^{-\frac{1}{r}}
\frac{(q^{2r+1}z_1/z;q^{2r})_\infty
(q^{2r+1}z_2/z;q^{2r})_\infty}{
(q^{2r-1}z_1/z;q^{2r})_\infty 
(q^{2r-1}z_2/z;q^{2r})_\infty}
F_k(q^{1-\frac{2s}{N}}z)
F_j(q^{1-\frac{2s}{N}}z_1)
F_j(q^{1-\frac{2s}{N}}z_2)
\right\}\nonumber\\
&\times&
z_1^{\frac{1}{r}}
\frac{(q^{2r-2}z_2/z_1;q^{2r})_\infty}
{(q^{2r+2}z_2/z_1;q^{2r})_\infty}
+(z_1\leftrightarrow z_2)=0.
\end{eqnarray}

\subsection{Free field realization}

In this section we give the free field realization
of the elliptic quantum group
$U_{q.p}(\widehat{sl_N})$ \cite{JKOS, KojimaKonno, AJMP}.
In what follows we restrict our interest to
level $c=1$.
Let us introduce the bosons 
$\beta_m^j, (1\leqq j \leqq N;m \in {\mathbb Z})$ by
\begin{eqnarray}
~[\beta_m^j,\beta_n^k]=
\left\{
\begin{array}{cc}
\displaystyle
m \frac{[(r-1)m]_q }{[rm]_q}
\frac{[(s-1)m]_q}{[sm]_q}\delta_{m+n,0}
&(1\leqq j=k \leqq N)\\
\displaystyle
-m q^{s m~{\rm sgn}(j-k)}
\frac{[(r-1)m]_q}{[rm]_q}
\frac{[m]_q}{[sm]_q}\delta_{m+n,0}
&(1\leqq j \neq k \leqq N).
\end{array}
\right.
\end{eqnarray}
We set the bosons $B_m^j$,
$(1\leqq j \leqq N; m \in {\mathbb Z}_{\neq 0})$ by
\begin{eqnarray}
B_m^j&=&(\beta_m^j-\beta_m^{j+1})q^{-jm},
~(1\leqq j \leqq N-1).%
%~B_m^N=(q^{-Nm}\beta_m^N-q^{(2s-N)m}\beta_m^1),\nonumber\\
\end{eqnarray}
They satisfy
\begin{eqnarray}
~[B_m^j,B_n^k]=m\frac{[(r-1)m]_q}{[rm]_q}
\frac{[{A}_{j,k}m]_q}{[m]_q}\delta_{m+n,0},~~
(1\leqq j,k \leqq N-1),
\end{eqnarray}
where $({A}_{j,k})_{1\leqq j,k \leqq N-1}$ 
is Cartan matrix of ${sl_N}$ type.
Let $\epsilon_\mu (1\leqq \mu \leqq N)$ be the orthonormal
basis of ${\mathbb R}^N$ with the inner
product $(\epsilon_\mu |\epsilon_\nu)=\delta_{\mu,\nu}$.
Let us set $\bar{\epsilon}_\mu=\epsilon_\mu-\epsilon$
where $\epsilon=\frac{1}{N}\sum_{\nu=1}^N \epsilon_\nu$.
Let $\alpha_\mu~(1\leqq \mu \leqq N-1)$ the simple root :
$\alpha_\mu=\bar{\epsilon}_\mu-\bar{\epsilon}_{\mu+1}$.
The type $sl_N$ weight lattice is the
linear span of $\bar{\epsilon}_\mu$,
$P=\sum_{\mu=1}^{N-1} {\mathbb Z}\bar{\epsilon}_\mu$.
Let us set
$P_\alpha, Q_\alpha$
$(\alpha \in P)$ by
\begin{eqnarray}
~[i P_\alpha,Q_\beta]=(\alpha|\beta),~~(\alpha,\beta \in P).
\end{eqnarray}
In what follows we 
deal with the bosonic Fock space 
${\cal F}_{l,k}$, generated by 
$\beta_{-m}^j (m>0)$ over the vacuum vector 
$|l,k \rangle$, where $l,k \in P$.
\begin{eqnarray}
&&{\cal F}_{l,k}={\mathbb C}[\{\beta_{-1}^j,
\beta_{-2}^j,\cdots\}_{1\leqq j \leqq N}]|l,k\rangle,~~
|l,k\rangle=
e^{i \sqrt{\frac{r}{r-1}}Q_l-
i \sqrt{\frac{r-1}{r}}Q_k}|0,0\rangle,\nonumber
\end{eqnarray}
where
\begin{eqnarray}
&&\beta_m^j |l,k\rangle=0,~(m>0),~~~
P_\alpha |l,k\rangle=\left(\alpha\left|
\sqrt{\frac{r}{r-1}}l-\sqrt{\frac{r-1}{r}}k
\right.\right)|l,k\rangle.\nonumber
\end{eqnarray}
Free field realizations of
$E_j(z), F_j(z), H_j^\pm(z)$ 
$(1\leqq j \leqq N-1)$ are given by
\begin{eqnarray}
E_j(z)&=&
e^{-i\sqrt{\frac{r}{r-1}}Q_{\alpha_j}}
(q^{(\frac{2s}{N}-1)j}z)
^{-\sqrt{\frac{r}{r-1}}P_{\alpha_j}+\frac{r}{r-1}}
\nonumber\\
&\times&
:\exp\left(
-\sum_{m \neq 0}\frac{1}{m}\frac{[rm]_q}{[(r-1)m]_q}
B_m^j (q^{(\frac{2s}{N}-1)j}z)^{-m}
\right):,\\
%%%%%%%%%%%%%%%%%%%%%%%%%%
F_j(z)&=&
e^{i \sqrt{\frac{r-1}{r}}Q_{\alpha_j}}
(q^{(\frac{2s}{N}-1)j}z)^{\sqrt{\frac{r-1}{r}}P_{\alpha_j}
+\frac{r-1}{r}}\nonumber\\
&\times&
:\exp\left(\sum_{m \neq 0}\frac{1}{m}B_m^j 
(q^{(\frac{2s}{N}-1)j}z)^{-m}
\right):,\\
%%%%%%%%%%%%%%%%%%%%%%%%%%
H_j^+(q^{\frac{1}{2}-r}z)&=&
q^{(1-\frac{2s}{N})2j}
e^{-\frac{i}{\sqrt{r (r-1)}}Q_{\alpha_j}}
(q^{(\frac{2s}{N}-1)j}z)^{-\frac{1}{\sqrt{r (r-1)}}P_{\alpha_j}
+\frac{1}{r (r-1)}}\nonumber\\
&\times&:\exp\left(
-\sum_{m \neq 0}\frac{1}{m}\frac{[m]_q}{[(r-1)m]_q}
B_m^j (q^{(\frac{2s}{N}-1)j}z)^{-m}\right):.
\end{eqnarray}
The free field realization for general level $c$ 
\cite{Kojima1}
is completely
different from those for level $c=1$.

\section{Commuting operators}

In this section
we construct two classes of infinitely many
commuting operators ${\cal G}_m$ \cite{KojimaShiraishi1} 
and $T_B(z)$ \cite{Kojima3}.

\subsection{Extended currents $E_N(z), F_N(z)$}

In this section we introduce
the extended currents 
${E}_N(z),{F}_N(z)$ \cite{KojimaShiraishi1}.
Let us set the extended current ${E}_N(z),{F}_N(z)$
by the similar commutation relations as 
the elliptic quantum group.
The extended currents $E_N(z),F_N(z)$ satisfy the following
commutation relations. 
\begin{eqnarray}
E_j(z_1){E}_{j+1}(z_2)
&=&
\frac{
\left[u_2-u_1+\frac{s}{N}\right]^*}
{\left[u_1-u_2+1-\frac{s}{N}\right]^*}
{E}_{j+1}(z_2){E}_j(z_1),~(j \in {\mathbb Z}/N {\mathbb Z}),
\nonumber\\
{E}_j(z_1)
{E}_j(z_2)&=&
\frac{[u_1-u_2+1]^*}
{[u_1-u_2-1]^*}
{E}_j(z_2){E}_j(z_1),
~(j \in {\mathbb Z}/N {\mathbb Z}),
\nonumber\\
{F}_j(z_1){F}_{j+1}(z_2)
&=&\frac{\left[u_2-u_1+\frac{s}{N}-1\right]}{
\left[u_1-u_2-\frac{s}{N}\right]
}{F}_{j+1}(z_2){F}_j(z_1),~(j \in {\mathbb Z}/N {\mathbb Z}),
\nonumber\\
{F}_j(z_1){F}_j(z_2)&=&
\frac{[u_1-u_2-1]}{[u_1-u_2+1]
}{F}_j(z_2){F}_j(z_1),~(j \in {\mathbb Z}/N {\mathbb Z}),
\nonumber\\
~[{E}_j(z_1),{F}_k(z_2)]
&=&
\frac{\delta_{j,k}}{q-q^{-1}}
\left(\delta(q^{-1}z_1/z_2)
{H}_j^+\left(q^{\frac{1}{2}}z_2\right)
-
\delta(qz_1/z_2)
{H}_j^-\left(q^{-\frac{1}{2}}z_2\right)\right),\nonumber\\
&&~~~~~~~~~~~~~~~~~~~~~
~~~~~~~~~~~~~~~~~~
(j,k \in {\mathbb Z}/N{\mathbb Z}),\nonumber
\end{eqnarray}
and other defining relations of the elliptic quantum group,
in which the suffix $j,k$ should be understood as $mod.~N$.
Free field realizations of the extended currents 
${E}_N(z), {F}_N(z)$ and
${H}_N^+(q^{\frac{1}{2}-r}z)={H}_N^-(q^{\frac{1}{2}-r}z)$
are given by
\begin{eqnarray}
{E}_N(z)&=&
e^{-i\sqrt{\frac{r}{r-1}}Q_{\alpha_N}}
(q^{2s-N}z)^{-\sqrt{\frac{r}{r-1}}P_{\bar{\epsilon}_N}+\frac{r}{2(r-1)}}
z^{\sqrt{\frac{r}{r-1}}P_{\bar{\epsilon}_1}+\frac{r}{2(r-1)}}
\nonumber\\
&\times&
:\exp\left(
-\sum_{m \neq 0}\frac{1}{m}\frac{[rm]_q}{[(r-1)m]_q}
B_m^N (q^{2s-N}z)^{-m}\right):,\\
%%%%%%%%%%%%%%%%%%%%%%%%%%%%%%%
{F}_N(z)&=&
e^{i\sqrt{\frac{r-1}{r}}Q_{\alpha_N}}
(q^{2s-N}z)^{\sqrt{\frac{r-1}{r}}P_{\bar{\epsilon}_N}+\frac{r-1}{2r}}
z^{-\sqrt{\frac{r-1}{r}}P_{\bar{\epsilon}_1}+\frac{r-1}{2r}}\nonumber\\
&\times&
:\exp\left(
-\sum_{m \neq 0}\frac{1}{m}
B_m^N (q^{2s-N}z)^{-m}\right):,\\
{H}_N^+(q^{\frac{1}{2}-r}z)&=&
q^{2(N-2s)}e^{-\frac{i}{\sqrt{r r^*}}Q_{\alpha_N}}
(q^{2s-N}z)^{-\frac{1}{\sqrt{r r^*}}
P_{\bar{\epsilon}_N}+\frac{1}{2r r^*}}
z^{\frac{1}{\sqrt{r r^*}}
P_{\bar{\epsilon}_1}+\frac{1}{2r r^*}}
\nonumber\\
&\times&
:\exp\left(
-\sum_{m \neq 0}\frac{1}{m}\frac{[m]_q}{[(r-1)m]_q}
B_m^N (q^{2s-N}z)^{-m}\right):.
\end{eqnarray}

\subsection{Extended currents $V(z), U(z)$}

In this section
we introduce the extended currents $V(z), U(z)$
\cite{AJMP, FKQ}.
In this section we consider the case $s=N$.
For our purpose it is convenient to introduce
\begin{eqnarray}
\overline{E_j}(z)=E_j(q^{-j}z),~~\overline{F}_j(z)=
F_j(q^{-j}z),~~(1\leqq j \leqq N-1).\nonumber
\end{eqnarray}
The extended currents $U(z), V(z)$ are given by
the following commutation relations.
\begin{eqnarray}
~\left[u_1-u_2+\frac{1}{2}\right]^*
V(z_1)\overline{E}_1(z_2)
&=&\left[u_2-u_1+\frac{1}{2}\right]^*
\overline{E}_1(z_2)V(z_1),\\
\overline{E}_{j}(z_1)V(z_2)&=&V(z_2)
\overline{E}_{j}(z_1)~~(2\leqq j \leqq N),
\\
~\left[u_1-u_2-\frac{1}{2}\right]
U(z_1)\overline{F}_1(z_2)
&=&\left[u_2-u_1-\frac{1}{2}\right]
\overline{F}_1(z_2)U(z_1),\\
\overline{F}_{j}(z_1)U(z_2)&=&
U(z_2)\overline{F}_{j}(z_1)~~(2\leqq j \leqq N).
\end{eqnarray}
\begin{eqnarray}
U(z_1)U(z_2)&=&
(z_1/z_2)^{\frac{r-1}{r}\frac{N-1}{N}}
\frac{\rho(z_2/z_1)}{\rho(z_1/z_2)}
U(z_2)U(z_1),\\
V(z_1)V(z_2)&=&
(z_1/z_2)^{-\frac{r}{r-1}\frac{N-1}{N}}
\frac{\rho^*(z_2/z_1)}{\rho^*(z_1/z_2)}
V(z_2)V(z_1),\\
U(z_1)V(z_2)&=&
z^{-\frac{N-1}{N}}
\frac{\Theta_{q^{2N}}(-qz)}{
\Theta_{q^{2N}}(-qz^{-1})}V(z_2)U(z_1),
\end{eqnarray}
where we set
\begin{eqnarray}
\rho(z)&=&
\frac{
(q^2z;q^{2r},q^{2N})_\infty 
(q^{2N+2r-2}z;q^{2r},q^{2N})_\infty}{
(q^{2r}z;q^{2r},q^{2N})_\infty
(q^{2N}z;q^{2r},q^{2N})_\infty},
\\
\rho^*(z)&=&
\frac{
(z;q^{2r^*},q^{2N})_\infty
(q^{2N+2r-2}z;q^{2r^*},q^{2N})_\infty}{
(q^{2r}z;q^{2r^*},q^{2N})_\infty
(q^{2N-2}z;q^{2r^*},q^{2N})_\infty}.
\end{eqnarray}
%The most large difference between 
%two currents ${F}_N(z)$
%and $U(z)$ is
%the commutation relation with $F_{N-1}(z)$.
%\begin{eqnarray}
%\left[u_1-u_2-\frac{1}{2}\right]{F}_{N-1}(z)
%{F}_{N}(z_2)
%&=&{F}_{N}(z_2)
%{F}_{N-1}(z_1)
%\left[u_2-u_1+\frac{1}{2}\right],\nonumber\\
%F_{N-1}(z_1)U(z_2)&=&U(z_2)F_{N-1}(z_1).\nonumber
%\end{eqnarray}
The free field realizations of
$U(z), V(z)$ are given by
\begin{eqnarray}
U(z)&=&
z^{\frac{r-1}{2r}\frac{N-1}{N}}
e^{-i\sqrt{\frac{r-1}{r}}Q_{\bar{\epsilon}_1}}
z^{-\sqrt{\frac{r-1}{r}}P_{\bar{\epsilon}_1}}
:\exp\left(-\sum_{m \neq 0}\frac{1}{m}\beta_m^1 z^{-m}
\right):,
\\
V(z)&=&
z^{\frac{r}{2(r-1)}\frac{N-1}{N}}
e^{i\sqrt{\frac{r}{r-1}}Q_{\bar{\epsilon}_1}}
z^{\sqrt{\frac{r}{r-1}}P_{\bar{\epsilon}_1}}
\nonumber\\
&\times&
:\exp\left(\sum_{m \neq 0}\frac{1}{m}\frac{[rm]_q}{[(r-1)m]_q}
\beta_m^1 (-z)^{-m}
\right):.
\end{eqnarray}

\subsection{Integral of motion}

In this section
we give a class of infinitely many commuting
operators ${\cal G}_m$,
$(m \in {\mathbb N})$ that we call the integral of motion
\cite{KojimaShiraishi1}.
In this section we consider the case $0<{\rm Re}(s)<N$.
Let us set the integral of motion ${\cal G}_m$,
$(m \in {\mathbb N})$ by integral
of the currents.
\begin{eqnarray}
{\cal G}_m
&=&\int \cdots \int
\prod_{t=1}^N \prod_{j=1}^m
\frac{dz_j^{(t)}}{z_j^{(t)}}
{F}_1(z_1^{(1)})
{F}_1(z_2^{(1)})
\cdots 
{F}_1(z_m^{(1)})\nonumber\\
&\times&
{F}_2(z_1^{(2)})
{F}_2(z_2^{(2)})\cdots 
{F}_2(z_m^{(2)})
\cdots
{F}_N(z_1^{(N)})
{F}_N(z_2^{(N)})\cdots 
{F}_N(z_m^{(N)})
\nonumber\\
&\times&
\frac{
\displaystyle
\prod_{t=1}^N
\prod_{1\leqq j<k \leqq m}
\left[u_j^{(t)}-u_k^{(t)}\right]
\left[u_k^{(t)}-u_j^{(t)}-1\right]
}{
\displaystyle
\prod_{t=1}^{N-1}
\prod_{j,k=1}^m
\left[u_j^{(t)}-u_k^{(t+1)}+1-\frac{s}{N}\right]
\prod_{j,k=1}^m
\left[u_j^{(1)}-u_k^{(N)}+\frac{s}{N}\right]}
\nonumber\\
&\times&\prod_{t=1}^N
\left[\sum_{j=1}^m(u_j^{(t)}-u_j^{(t+1)})-
\sqrt{r r^*}P_{\bar{\epsilon}_{t+1}}
\right].
\end{eqnarray}
Here we set $z_j^{(t)}=q^{2u_j^{(t)}}$.
Here the integral contour
encircles $z_j^{(t)}=0$, 
$(1\leqq t\leqq N; 1\leqq j \leqq m)$
in such a way that 
\begin{eqnarray}
&&|q^{\frac{2s}{N}+2lr}z_k^{(t+1)}|
<|z_j^{(t)}|<
|q^{-2+\frac{2s}{N}-2lr}z_k^{(t+1)}|,~(1\leqq t \leqq N-1),
\nonumber\\
&&|q^{2-\frac{2s}{N}+2lr}z_k^{(1)}|
<|z_j^{(N)}|<
|q^{-\frac{2s}{N}-2lr}z_k^{(1)}|,\nonumber
\end{eqnarray}
for $1\leqq j, k \leqq m$ and $l \in {\mathbb N}$.
Let us set the integral of motion
${\cal G}_m^*$, $(m \in {\mathbb N})$ as similar way.
\begin{eqnarray}
{\cal G}_m^*
&=&\int \cdots \int
\prod_{t=1}^N \prod_{j=1}^m
\frac{dz_j^{(t)}}{z_j^{(t)}}
E_1(z_1^{(1)})E_1(z_2^{(1)})
\cdots E_1(z_m^{(1)})\nonumber\\
&\times&
E_2(z_1^{(2)})E_2(z_2^{(2)})\cdots E_2(z_m^{(2)})
\cdots
E_N(z_1^{(N)})E_N(z_2^{(N)})\cdots E_N(z_m^{(N)})
\nonumber\\
&\times&
\frac{
\displaystyle
\prod_{t=1}^N
\prod_{1\leqq j<k \leqq m}
\left[u_j^{(t)}-u_k^{(t)}\right]^*
\left[u_k^{(t)}-u_j^{(t)}+1\right]^*
}{
\displaystyle
\prod_{t=1}^{N-1}
\prod_{j,k=1}^m
\left[u_j^{(t)}-u_k^{(t+1)}-\frac{s}{N}\right]^*
\prod_{j,k=1}^m
\left[u_j^{(1)}-u_k^{(N)}-1+\frac{s}{N}\right]^*
}\nonumber
\\
&\times&\prod_{t=1}^N
\left[\sum_{j=1}^m(u_j^{(t)}-u_j^{(t+1)})-
\sqrt{r r^*}P_{\bar{\epsilon}_{t+1}}
\right]^*.
\end{eqnarray}
Here the integral contour
encircles $z_j^{(t)}=0$, $(1\leqq t\leqq N; 1\leqq j \leqq m)$
in such a way that 
\begin{eqnarray}
&&|q^{-2+\frac{2s}{N}+2lr^*}z_k^{(t+1)}|
<|z_j^{(t)}|<
|q^{\frac{2s}{N}-2lr^*}z_k^{(t+1)}|,~(1\leqq t \leqq N-1),
\nonumber\\
&&|q^{-\frac{2s}{N}+2lr^*}z_k^{(1)}|
<|z_j^{(N)}|<
|q^{2-\frac{2s}{N}-2lr^*}z_k^{(1)}|,\nonumber
\end{eqnarray}
for $1\leqq j, k \leqq m$ and $l \in {\mathbb N}$.

~\\

The integral of motion
${\cal G}_m$ and ${\cal G}_m^*$
commute with each other.
\begin{eqnarray}
~[{\cal G}_m,{\cal G}_n]=0,~~
~[{\cal G}_m^*,{\cal G}_n^*]=0,~~
~[{\cal G}_m,{\cal G}_n^*]=0,~~(m,n \in {\mathbb N}).
\end{eqnarray}
These commutation relations are shown by considering
the Feigin-Odesskii algebra \cite{FO}.
When we take the limit $r \to \infty$,
our integral of motion ${\cal G}_m$ becomes
those of conformal field theory
\cite{BLZ, Kojima2}.
In the limit $r \to \infty$,
the theta functions in integrand disappear,
hence we know that elliptic deformation
is nontrivial.
The integral of motion of $U_{q,p}(\widehat{sl_2})$
in general level $c$ is constructed in \cite{KojimaShiraishi2}.

\subsection{Vertex operator}

In this section we introduce the vertex operator 
$\Phi^{(a,b)}(z)$ that plays an essential role
in construction of the boundary transfer matrix $T_B(z)$.
In this section we consider the case $r \geqq N+2,
(r \in {\mathbb N})$ and $s=N$. 
Let's recall $sl_N$ weight lattice 
$P=\sum_{\mu=1}^{N-1} {\mathbb Z}\bar{\epsilon}_\mu$
introduced in previous section.
Let $\omega_\mu~(1\leqq \mu \leqq N-1)$ be 
the fundamental weights, which satisfy
\begin{eqnarray}
(\alpha_\mu|\omega_\nu)=\delta_{\mu,\nu},
~~(1\leqq \mu,\nu \leqq N-1).\nonumber
\end{eqnarray}
Explicitly we set
$\omega_\mu=\sum_{\nu=1}^\mu \bar{\epsilon}_\nu.$
For $a \in P$ we set $a_\mu$ and $a_{\mu,\nu}$ by
\begin{eqnarray}
a_{\mu,\nu}=a_{\mu}-a_{\nu},~~~
a_{\mu}=(a+\rho|\bar{\epsilon}_\mu),~~~(\mu, \nu \in P).
\nonumber
\end{eqnarray}
Here we set
$\rho=\sum_{\mu=1}^{N-1}\omega_\mu$.
Let us set the restricted path $P_{r-N}^+$ by
\begin{eqnarray}
P_{r-N}^+=\{a=\sum_{\mu=1}^{N-1}c_\mu \omega_\mu \in P|
c_\mu \in {\mathbb Z}, c_\mu \geqq 0, 
\sum_{\mu=1}^{N-1}c_\mu \leqq r-N \}.
\nonumber
\end{eqnarray}
For $a \in P_{r-N}^+$, condition 
$0<a_{\mu,\nu}<r,~(1\leqq \mu<\nu\leqq N-1)$ holds.
We recall elliptic solutions of
the Yang-Baxter equation of face type.
An ordered pair $(b,a)\in P^2$ is called
admissible if and only if there exists
$\mu~(1\leq \mu \leq N)$ such that
$b-a=\bar{\epsilon}_\mu$.
An ordered set of four weights $(a,b,c,d)\in P^4$
is called an admissible configuration
around a face if and only if
the ordered pairs $(b,a)$, $(c,b)$, $(d,a)$ and $(c,d)$
are admissible.
Let us set
the Boltzmann weight functions 
$W\left(\left.\begin{array}{cc}
c&d\\
b&a
\end{array}\right|u\right)$
associated with admissible configuration
$(a,b,c,d)\in P^4$ \cite{JMO}.
For $a \in P_{r-N}^+$ and $\mu \neq \nu$, we set
\begin{eqnarray}
&&W\left(\left.
\begin{array}{cc}
a+2\bar{\epsilon}_\mu & a+\bar{\epsilon}_\mu\\
a+\bar{\epsilon}_\mu & a
\end{array}\right|u\right)=R(u),\label{def:B1}\\
&&W\left(\left.
\begin{array}{cc}
a+\bar{\epsilon}_\mu+\bar{\epsilon}_\nu & 
a+\bar{\epsilon}_\mu\\
a+\bar{\epsilon}_\nu & a
\end{array}\right|u\right)=R(u)\frac{[u][a_{\mu,\nu}-1]}
{[u-1][a_{\mu,\nu}]},
\label{def:B2}\\
&&W\left(\left.
\begin{array}{cc}
a+\bar{\epsilon}_\mu+\bar{\epsilon}_\nu 
& a+\bar{\epsilon}_\nu\\
a+\bar{\epsilon}_\nu & a
\end{array}\right|u\right)=
R(u)\frac{[u-a_{\mu,\nu}][1]}{
[u-1][a_{\mu,\nu}]}.
\label{def:B3}
\end{eqnarray}
The normalizing function $R(u)$ is 
given by 
\begin{eqnarray}
R(u)&=&z^{\frac{r-1}{r}\frac{N-1}{N}}
\frac{\varphi(z^{-1})}{\varphi(z)},~~~
\varphi(z)=\frac{
(q^{2}z;q^{2r},q^{2N})_\infty 
(q^{2r+2N-2}z;q^{2r},q^{2N})_\infty}{
(q^{2r}z;q^{2r},q^{2N})_\infty 
(q^{2N}z;q^{2r},q^{2N})_\infty}.
\nonumber\\
\end{eqnarray}
Because $0<a_{\mu,\nu}<r~(1\leqq \mu<\nu \leqq N-1)$
holds for $a \in P_{r-N}^+$,
the Boltzmann weight functions
are well defined.
The Boltzmann weight functions 
satisfy the Yang-Baxter equation of the face type.
\begin{eqnarray}
&&\sum_{g}
W\left(\left.\begin{array}{cc}
d&e\\
c&g
\end{array}
\right|u_1\right)
W\left(\left.\begin{array}{cc}
c&g\\
b&a
\end{array}
\right|u_2\right)
W\left(\left.\begin{array}{cc}
e&f\\
g&a
\end{array}
\right|u_1-u_2\right)
\nonumber\\
&=&
\sum_{g}
W\left(\left.\begin{array}{cc}
g&f\\
b&a
\end{array}
\right|u_1\right)
W\left(\left.\begin{array}{cc}
d&e\\
g&f
\end{array}
\right|u_2\right)
W\left(\left.\begin{array}{cc}
d&g\\
c&b
\end{array}
\right|u_1-u_2\right).
\label{eqn:Boltzmann1}
\end{eqnarray}
We set the normalization function 
$\varphi(z)$ such that the minimal eigenvalue of 
the corner transfer matrix becomes 1
\cite{Baxter}.
The vertex operator $\Phi^{(b,a)}(z)$ 
and the dual vertex operator $\Phi^{*(a,b)}(z)$
associated with the elliptic quantum group
$U_{q,p}(\widehat{sl_N})$,
are the operators
which satisfy the following commutation relations,
\begin{eqnarray}
\Phi^{(a,b)}(z_1)
\Phi^{(b,c)}(z_2)
&=&\sum_{g}
W\left(\left.\begin{array}{cc}
a&g\\
b&c
\end{array}
\right|u_2-u_1\right)
\Phi^{(a,g)}(z_2)
\Phi^{(g,c)}(z_1),\label{eqn:VO1}\\
%%%%%%%%%%%%%%%%%%%%%%%%%%%%%%%%%%%%
\Phi^{(a,b)}(z_1)
\Phi^{*(b,c)}(z_2)
&=&
\sum_{g}
W\left(\left.\begin{array}{cc}
g&c\\
a&b
\end{array}
\right|u_1-u_2\right)
\Phi^{*(a,g)}(z_2)
\Phi^{(g,c)}(z_1),\label{eqn:VO2}\\
%%%%%%%%%%%%%%%%%%%%%%%%%%%%%%%%%%%%
\Phi^{*(a,b)}(z_1)
\Phi^{*(b,c)}(z_2)
&=&
\sum_{g}
W\left(\left.\begin{array}{cc}
c&b\\
g&a
\end{array}
\right|u_2-u_1\right)
\Phi^{*(a,g)}(z_2)
\Phi^{*(g,c)}(z_1).\nonumber\\
\label{eqn:VO3}
\end{eqnarray}
and the inversion relation,
\begin{eqnarray}
\Phi^{(a,g)}(z)\Phi^{*(g,b)}(z)=\delta_{a,b}.
\label{eqn:inversion1}
\end{eqnarray}
We give free field realization of the vertex operator.
In what follows we set
$l=b+\rho, k=a+\rho$, $(a \in P_{r-N}^+, b \in P_{r-N-1}^+)$ 
and $\pi_\mu=\sqrt{r(r-1)}P_{\bar{\epsilon}_\mu},~
\pi_{\mu, \nu}=\pi_\mu-\pi_\nu$.
We give the free field realization of
the vertex operators
$\Phi^{(a+\bar{\epsilon}_{\mu},a)}(z)$, 
$(1\leqq \mu \leqq N-1)$ \cite{AJMP} by
\begin{eqnarray}
\Phi^{(a+\bar{\epsilon}_1,a)}
(z_0^{-1})&=&
U(z_0),\nonumber
\\
\Phi^{(a+\bar{\epsilon}_\mu,a)}(z_0^{-1})
&=&
\oint
\cdots \oint 
\prod_{j=1}^{\mu-1}
\frac{dz_j}{2\pi i z_j} U(z_0)
\overline{F}_1(z_1)
\overline{F}_2(z_2)
\cdots
\overline{F}_{\mu-1}(z_{\mu-1})\nonumber\\
&\times&
\prod_{j=1}^{\mu-1}
\frac{[u_j-u_{j-1}+\frac{1}{2}-\pi_{j,\mu}]}
{[u_j-u_{j-1}-\frac{1}{2}]}.
\end{eqnarray}
Here we set $z_j=q^{2u_j}$.
We take the integration contour to be simple closed 
curve that encircles 
$z_j=0, q^{1+2rs}z_{j-1}, (s \in {\mathbb N})$
but not
$z_j=q^{-1-2rs}z_{j-1}, (s \in {\mathbb N})$
for $1\leq j \leq \mu-1$.
The $\Phi^{(a+\bar{\epsilon}_\mu,a)}(z)$ 
is an operator such that
$\Phi^{(a+\bar{\epsilon}_\mu,a)}(z):
{\cal F}_{l,k}\to {\cal F}_{l,k+\bar{\epsilon}_\mu}$.
The free field realization of the dual vertex operator
$\Phi^{*(a,b)}(z)$
is given by similar way \cite{AJMP}.
The vertex operator $\Phi^{(a,b)}(z)$ plays an important role
in construction of the correlation functions
of the $U_{q,p}(\widehat{sl_N})$ face model
\cite{AJMP,LP}.

\subsection{Boundary transfer matrix}

In this section
we introduce
the boundary transfer matrix $T_B(z)$
\cite{Kojima3},
following theory of boundary Yang-Baxter equation
\cite{S, JKKKM}.
In this section we consider the case $r \geqq N+2,
(r \in {\mathbb N})$ and $s=N$. 
An order set of three weights $(a,b,g)
\in P^3$
is called 
an admissible configuration at a boundary
if and only if
the ordered pairs $(g,a)$ and $(g,b)$ are admissible. 
Let us set the boundary Boltzmann weight functions
$
K\left(
\left.\begin{array}{cc}
&a\\
g&\\
&b
\end{array}
\right|u\right)
$ for admissible weights $(a,b,g)$ as following \cite{BFKZ}.
\begin{eqnarray}
K\left(
\left.\begin{array}{cc}
&a\\
a+\bar{\epsilon}_\mu&\\
&b
\end{array}
\right|u\right)=
z^{\frac{r-1}{r}\frac{N-1}{N}-\frac{2}{r}
a_1}\frac{h(z)}{h(z^{-1})}
\frac{[c-u][a_{1,\mu}+c+u]}
{[c+u][a_{1,\mu}+c-u]}
\delta_{a,b}.\nonumber\\
\end{eqnarray}
In this paper, we consider
the case of continuous parameter $0<c<1$.
The normalization function
$h(z)$ is given by following \cite{Kojima3}.
\begin{eqnarray}
h(z)&=&
\frac{
(q^{2r+2N-2}/z^2;q^{2r},q^{4N})_\infty 
(q^{2N+2}/z^2;q^{2r},q^{4N})_\infty}{
(q^{2r}/z^2;q^{2r},q^{4N})_\infty 
(q^{4N}/z^2;q^{2r},q^{4N})_\infty}\nonumber\\
&\times&
\frac{
(q^{2N+2c}/z;q^{2r},q^{2N})_\infty
(q^{2r-2c}/z;q^{2r},q^{2N})_\infty}{
(q^{2N+2r-2c-2}/z;q^{2r},q^{2N})_\infty
(q^{2c+2}/z;q^{2r},q^{2N})_\infty}\\
&\times&
\prod_{j=2}^N 
\frac{
(q^{2r+2N-2c-2a_{1,j}}/z;q^{2r},q^{2N})_\infty
(q^{2c+2a_{1,j}}/z;q^{2r},q^{2N})_\infty}{
(q^{2r+2N-2c-2a_{1,j}-2}/z;q^{2r},q^{2N})_\infty
(q^{2c+2+2a_{1,j}}/z;q^{2r},q^{2N})_\infty}.\nonumber
\end{eqnarray}
The boundary Boltzmann weight functions 
and the Boltzmann weight functions
satisfy the boundary Yang-Baxter equation \cite{S}.
\begin{eqnarray}
&&\sum_{f,g}
W\left(\left.\begin{array}{cc}
c&f\\
b&a
\end{array}\right|u_1-u_2\right)
W\left(\left.\begin{array}{cc}
c&d\\
f&g
\end{array}
\right|u_1+u_2\right)
K\left(\left.\begin{array}{cc}
~&g\\
f&\\
~&a
\end{array}
\right|u_1\right)
K\left(\left.\begin{array}{cc}
~&e\\
d&\\
~&g
\end{array}
\right|u_2\right)
\nonumber\\
&=&
\sum_{f,g}
W\left(\left.\begin{array}{cc}
c&d\\
f&e
\end{array}\right|u_1-u_2\right)
W\left(\left.\begin{array}{cc}
c&f\\
b&g
\end{array}
\right|u_1+u_2\right)
K\left(\left.\begin{array}{cc}
~&e\\
f&\\
~&g
\end{array}
\right|u_1\right)
K\left(\left.\begin{array}{cc}
~&g\\
b&\\
~&a
\end{array}
\right|u_2\right).\nonumber
\\
\label{eqn:boundaryYBE2}
\end{eqnarray}
We set the normalization function $h(z)$ such that
the minimal eigenvalue of the boundary transfer matrix $T_B(z)$
becomes $1$.
We define the boundary transfer matrix $T_B(z)$
for the elliptic quantum group $U_{q,p}(\widehat{sl_N})$.
\begin{eqnarray}
T_B(z)=\sum_{\mu=1}^N
\Phi^{*(a,a+\bar{\epsilon}_\mu)}(z^{-1})
K\left(\left.
\begin{array}{cc}
~& a\\
a+\bar{\epsilon}_\mu &\\
~& a
\end{array}
\right|u\right)
\Phi^{(a+\bar{\epsilon}_\mu,a)}(z).
\label{def:boundary-transfer}
\end{eqnarray}
The boundary $T_B(z)$ commute with each other.
\begin{eqnarray}
~[T_B(z_1),T_B(z_2)]=0,~~~{\rm for~any~}z_1, z_2.
\end{eqnarray}
This commutativity is
consequence of
the commutation relations 
of the vertex operators (\ref{eqn:VO1}),
(\ref{eqn:VO2}), (\ref{eqn:VO3}), 
and boundary Yang-Baxter equation
(\ref{eqn:boundaryYBE2}).

\section{Diagonalization}

In this section we diagonalize
the boundary transfer matrix $T_B(z)$,
using free field realization of the vertex operators
\cite{Kojima3, AJMP, FKQ}.
In this section we consider the case $r \geqq N+2$,
$(r \in {\mathbb N})$ and $s=N$.

\subsection{Boundary state}

We call the eigenvector 
$|B\rangle$ with the eigenvalue $1$
the boundary state.
\begin{eqnarray}
T_B(z)|B\rangle=|B\rangle.
\end{eqnarray}
We construct
the free field realization of 
the boundary state $|B\rangle$,
analyzing those of the
transfer matrix $T_B(z)$.
The free field realization of 
the boundary state $|B\rangle$
is given as following \cite{Kojima3}.
\begin{eqnarray}
|B\rangle=e^F|k,k\rangle.
\end{eqnarray}
Here we have set
\begin{eqnarray}
F&=&
-\frac{1}{2}\sum_{m>0}
\sum_{j=1}^{N-1}\sum_{k=1}^{N-1}
\frac{1}{m}\frac{[rm]_q}{[(r-1)m]_q}
I_{j,k}(m)B_{-m}^j B_{-m}^k
+\sum_{m>0}\sum_{j=1}^{N-1}\frac{1}{m}
D_j(m)\beta_{-m}^j,\nonumber\\
\end{eqnarray}
where
\begin{eqnarray}
D_j(m)&=&
-\theta_m\left(\frac{[(N-j)m/2]_q[rm/2]_q^+
q^{\frac{(3j-N-1)m}{2}}}
{[(r-1)m/2]_q}\right)\nonumber\\
&&+\frac{q^{(j-1)m}[(-r+2\pi_{1,j}+2c-j+2)m]_q}{[(r-1)m]_q}
\nonumber\\
&&+\frac{[m]_q 
q^{(r-2c+2j-2)m}}{[(r-1)m]_q}
\left(\sum_{k=j+1}^{N-1}q^{-2m \pi_{1,k}}\right)\nonumber\\
&&+\frac{q^{(2j-N)m}[(r-2\pi_{1,N}-2c+N-1)m]_q}{[(r-1)m]_q},
\end{eqnarray}
\begin{eqnarray}
I_{j,k}(m)=\frac{[jm]_q[(N-k)m]_q}{[m]_q[Nm]_q}
=I_{k,j}(m)~~(1\leqq j\leqq k \leqq N-1).
\end{eqnarray}
Here we have used
$[a]_q^+=q^a+q^{-a}$
and $\theta_m(x)=
\left\{\begin{array}{cc}
x,& m:~{\rm even}\\
0,& m:~{\rm odd}
\end{array}\right.$.

\subsection{Excited states}

In this section we construct diagonalization
of the boundary transfer matrix $T_B(z)$
by using the boundary state $|B\rangle$
and type-II vertex operator $\Psi^{*(b,a)}(z)$. 
Let us introduce
type-II vertex operator
$\Psi^{*(b,a)}(z)$ \cite{FKQ} by
the following commutation relations,
\begin{eqnarray}
\Psi^{*(a,b)}(z_1)
\Psi^{*(b,c)}(z_2)
&=&\sum_{g}
W^*\left(\left.\begin{array}{cc}
a&g\\
b&c
\end{array}
\right|u_1-u_2\right)
\Psi^{*(a,g)}(z_2)
\Psi^{*(g,c)}(z_1),\nonumber\\
\\
\Phi^{(d,c)}(z_1)\Psi^{*(b,a)}(z_2)
&=&
\chi(z_2/z_1)
\Psi^{* (b,a)}(z_2)\Phi^{(d,c)}(z_1),
\label{eqn:IIVO2}\\
\Phi^{*(c,d)}(z_1)\Psi^{*(b,a)}(z_2)
&=&
\chi(z_1/z_2)
\Psi^{* (b,a)}(z_2)\Phi^{*(c,d)}(z_1),
\label{eqn:IIVO3}
\end{eqnarray}
where we have set
$\chi(z)=z^{-\frac{N-1}{N}}
\frac{\Theta_{q^{2N}}(-qz)}{
\Theta_{q^{2N}}(-qz^{-1})}$
and 
$W^*\left(\left.\begin{array}{cc}
a&g\\
b&c
\end{array}
\right|u\right)$
is obtained by substitution $r \to r^*$ of
the Boltzmann weight functions
$W\left(\left.\begin{array}{cc}
a&g\\
b&c
\end{array}
\right|u\right)$ defined in
(\ref{def:B1}), (\ref{def:B2}), (\ref{def:B3}).
Let us set $l=b+\rho, k=a+\rho$, $(a \in P_{r-N}^+,
b \in P_{r-N-1}^+)$.
The free field realization
of the type-II vertex
operators $\Psi_\mu^{*(b,a)}(z)$, $(1\leqq \mu \leqq N-1)$
are give by
\begin{eqnarray}
\Psi^{*(b+\bar{\epsilon}_1,b)}
(z_0^{-1})&=&
V(z_0),\nonumber
\\
\Psi^{*(b+\bar{\epsilon}_\mu,b)}(z_0^{-1})
&=&
\oint
\cdots \oint 
\prod_{j=1}^{\mu-1}
\frac{dz_j}{2\pi i z_j} V(z_0)
\overline{E}_1(z_1)
\overline{E}_2(z_2)
\cdots
\overline{E}_{\mu-1}(z_{\mu-1})\nonumber\\
&\times&
\prod_{j=1}^{\mu-1}
\frac{[u_j-u_{j-1}-\frac{1}{2}+\pi_{j,\mu}]^*}
{[u_j-u_{j-1}+\frac{1}{2}]^*}.
\end{eqnarray}
We take the integration contour to be simple closed 
curve that encircles 
$z_j=0, q^{-1+2r^*s}z_{j-1}, (s \in {\mathbb N})$
but not
$z_j=q^{1-2r^*s}z_{j-1}, (s \in {\mathbb N})$
for $1\leq j \leq \mu-1$.
The $\Psi^{*(b+\bar{\epsilon}_\mu,b)}(z)$ 
is an operator such that
$\Psi^{*(b+\bar{\epsilon}_\mu,b)}(z):
{\cal F}_{l,k}\to {\cal F}_{l+\bar{\epsilon}_\mu,k}$.
We introduce 
the vectors $|\xi_1,\xi_2,\cdots,\xi_M
\rangle_{\mu_1,\mu_2,\cdots,\mu_M}$
$(1\leqq \mu_1,\mu_2,\cdots,\mu_M \leqq N)$. 
\begin{eqnarray}
&&|\xi_1,\xi_2,\cdots,\xi_M
\rangle_{\mu_1,\mu_2,\cdots,\mu_M}\nonumber\\
&=&
\Psi^{*(b+\bar{\epsilon}_{\mu_1}
+\bar{\epsilon}_{\mu_2}+\cdots
+\bar{\epsilon}_{\mu_M},
b+\bar{\epsilon}_{\mu_2}+\cdots
+\bar{\epsilon}_{\mu_M})}(\xi_1)
\times \cdots \nonumber\\
&&\times \cdots
\Psi^{*(b+\bar{\epsilon}_{\mu_{M-1}}
+\bar{\epsilon}_{\mu_M},b+\bar{\epsilon}_{\mu_M})}(\xi_{M-1})
\Psi^{*(b+\bar{\epsilon}_{\mu_M},b)}(\xi_M)
|B \rangle.
\end{eqnarray}
We construct many eigenvectors of $T_B(z)$.
\begin{eqnarray}
&&T_B(z)|\xi_1,\xi_2,\cdots,\xi_M
\rangle_{\mu_1,\mu_2,\cdots,\mu_M}\nonumber\\
&=&
\prod_{j=1}^M \chi(\xi_j/z)\chi(1/\xi_j z)~
|\xi_1,\xi_2,\cdots,\xi_M
\rangle_{\mu_1,\mu_2,\cdots,\mu_M}.
\end{eqnarray}
The vectors $
|\xi_1,\xi_2,\cdots,\xi_M
\rangle_{\mu_1,\mu_2,\cdots,\mu_M}$
are the basis of the space of the state of 
the boundary $U_{q,p}(\widehat{sl_N})$ face model
\cite{LP, MW, Kojima3}.
It is thought that our method can be extended to
more general elliptic quantum group $U_{q,p}(g)$.

~\\
{\large \bf Acknowledgements}\\
\\
The author would like to thank to
Branko Dragovich, Vladimir Dobrev,
Sergei Vernov, Paul Sorba,
Igor Salom, Dragan Savic and 
the organizing committee of 6th Mathematical 
Physics Meetings : Summer School and Conference
on Modern Mathematical Physics, held in Belgrade, Serbia.
This work is supported by
the Grant-in Aid
for Scientific Research {\bf C} (21540228)
from Japan Society for Promotion of Science.

\begin{appendix}

\end{appendix}


\begin{thebibliography}{99}
\bibitem{JM}M.Jimbo and T.Miwa,
{\it Algebraic Analysis of Solvable Lattice Models}
CBMS Regional Conference Series in Mathematics
{\bf 85} AMS 1994.
%%%%%%%%%%%%%%%%%%%%%
\bibitem{JKOS}M.Jimbo,H.Konno,S.Odake and J.Shiraishi,
{\it Commun.Math.Phys.}{\bf 199} (1999) 605.
%%%%%%%%%%%%%%%%%%%%%
\bibitem{KojimaKonno}T.Kojima and H.Konno,
{\it Commun.Math.Phys.}{\bf 239} (2003) 405.
%%%%%%%%%%%%%%%%%%%%
\bibitem{KojimaShiraishi1}T.Kojima and J.Shiraishi,
{\it Commun.Math.Phys.}{\bf 283} (2008) 795.
%%%%%%%%%%%%%%%%%%%%
\bibitem{AJMP}Y.Asai,M.Jimbo,T.Miwa and Ya.Pugai,
{\it J.Phys.}{\bf A29} (1996) 6595.
%%%%%%%%%%%%%%%%%%%%
\bibitem{Kojima3}T.Kojima, accepted for publication
in {\it J.Math.Phys.} (2010) [arXiv.1007.3795].
%%%%%%%%%%%%%%%%%%%%
\bibitem{FO}B.Feigin and A.Odesskii,
{\it Internat.Math.Res.Notices}{\bf 11} (1997) 531.
%%%%%%%%%%%%%%%%%%%%
\bibitem{S}E.Sklyanin, 
{\it J.Phys.}{\bf A21} (1988) 74.
%%%%%%%%%%%%%%%%%%%%
\bibitem{BLZ}V.Bazhanov, S.Lukyanov, Al.Zamolodchikov,
{\it Commun.Math.Phys.}{\bf 177} (1996) 381.
%%%%%%%%%%%%%%%%%%%%
\bibitem{Kojima2}T.Kojima, {\it J.Phys.}{\bf A41}
(2008) 355206 (16pp).
%%%%%%%%%%%%%%%%%%%%
\bibitem{LP}S.Lukyanov and Ya.Pugai,
{\it Nocl.Phys.}{\bf B473} (1996) 631.
%%%%%%%%%%%%%%%%%%%%
\bibitem{MW}T.Miwa and R.Weston, 
{\it Nucl.Phys.}{\bf B486} (1997) 517.
%%%%%%%%%%%%%%%%%%%%
\bibitem{FKQ}H.Furutsu, T.Kojima and Y.Quano,
{\it Int.J.Mod.Phys.}{\bf A15} (2000) 1533.
%%%%%%%%%%%%%%%%%%%%
\bibitem{Drinfeld}V.G.Drinfeld,
{\it Soviet.Math.Dokl.}{\bf 36} (1988) 212.
%%%%%%%%%%%%%%%%%%%%
\bibitem{BFKZ}M.T.Batchelor, V.Fridkin, A.Kuniba and Y.K.Zhou,
{\it Phys.Lett.}{\bf B276} (1996) 266.
%%%%%%%%%%%%%%%%%%%%
\bibitem{JKKKM}M.Jimbo,R.Kedem.T.Kojima,H.Konno and T.Miwa,
{\it Nucl.Phys.}{\bf B441} (1995) 437.
%%%%%%%%%%%%%%%%%%%%
\bibitem{Kojima1}T.Kojima, 
{\it Int.J.Mod.Phys.}{\bf A24} (2009) 5561.
%%%%%%%%%%%%%%%%%%%%
\bibitem{KojimaShiraishi2}T.Kojima and J.Shiraishi,
{\it J.Geometry, Integrability and Quantization}
{\bf X} (2009) 183.
%%%%%%%%%%%%%%%%%%%%
\bibitem{JMO}M.Jimbo, T.Miwa and S.Odake,
{\it Nucl.Phys.}{\bf B300} (1988) 507.
%%%%%%%%%%%%%%%%%%%%
\bibitem{FK}H.Furutsu and T.Kojima, {\it J.Math.Phys.}
{\bf 41} (2000) 4413.
%%%%%%%%%%%%%%%%%%%%
\bibitem{Baxter}R.Baxter,
{\it Exactly Solved Models in Statistical Mechanics}, 
Academic Press 1982.
\end{thebibliography}
\end{document}